\documentclass[twocolumn]{aastex61}
\pdfoutput=1 
\usepackage{amsmath,amstext}
\usepackage[T1]{fontenc}
\usepackage{apjfonts} 
\usepackage[figure,figure*]{hypcap}
\usepackage{subfigure}
\usepackage{xcolor}
\usepackage{soul}
\setstcolor{red}

\shorttitle{Discovery of the Optical Counterpart of GW170817 with DECam}
\shortauthors{Soares-Santos, et al.}

\begin{document}

\title{The Electromagnetic Counterpart of the Binary Neutron Star Merger LIGO/VIRGO GW170817. I. Discovery of  the  Optical Counterpart Using the Dark Energy Camera}


\author{M.~Soares-Santos}
\affiliation{Department of Physics, Brandeis University, Waltham, MA 02453, USA}
\affiliation{Fermi National Accelerator Laboratory, P. O. Box 500, Batavia, IL 60510, USA}
\author{D.~E.~Holz}
\affiliation{Enrico Fermi Institute, Department of Physics, Department of Astronomy and Astrophysics,\\and Kavli Institute for Cosmological Physics, University of Chicago, Chicago, IL 60637, USA}
\author{J.~Annis}
\affiliation{Fermi National Accelerator Laboratory, P. O. Box 500, Batavia, IL 60510, USA}
\author{R.~Chornock}
\affiliation{Astrophysical Institute, Department of Physics and Astronomy, 251B Clippinger Lab, Ohio University, Athens, OH 45701, USA}
\author{K.~Herner}
\affiliation{Fermi National Accelerator Laboratory, P. O. Box 500, Batavia, IL 60510, USA}
\author{E.~Berger}
\affiliation{Harvard-Smithsonian Center for Astrophysics, 60 Garden Street, Cambridge, Massachusetts 02138, USA}
\author{D.~Brout}
\affiliation{Department of Physics and Astronomy, University of Pennsylvania, Philadelphia, PA 19104, USA}
\author{H.~Chen}
\affiliation{Kavli Institute for Cosmological Physics, University of Chicago, Chicago, IL 60637, USA}
\author{R.~Kessler}
\affiliation{Kavli Institute for Cosmological Physics, University of Chicago, Chicago, IL 60637, USA}
\author{M.~Sako}
\affiliation{Department of Physics and Astronomy, University of Pennsylvania, Philadelphia, PA 19104, USA}
% ...

\collaboration{(The Dark Energy Survey and The Dark Energy Camera GW-EM Collaboration)}

\begin{abstract}
We present the Dark Energy Camera (DECam) discovery of the optical counterpart of the first binary neutron star merger detected through gravitational wave emission, GW170817.  Our observations commenced 10.5 hours post-merger, as soon as the localization region became accessible from Chile. We imaged 70 deg$^2$ in the $i$ and $z$ bands, covering 93\% of the initial integrated localization probability, to a depth necessary to identify likely optical counterparts (e.g., a kilonova).  At 11.4 hours post-merger we detected a bright optical transient located $10.6''$ from the nucleus of NGC\,4993 at redshift $z=0.0098$, consistent (for $H_0 = 70$\, km s$^{-1}$ Mpc$^{-1}$) with the distance of $40 \pm 8$\, Mpc reported by the  LIGO Scientific Collaboration and the Virgo Collaboration (LVC).  At detection the transient had magnitudes $i\approx 17.30$ and $z\approx 17.45$, and thus an absolute magnitude of $M_i = -15.7$, in the luminosity range expected for a kilonova. We identified 1,500 potential transient candidates. Applying simple selection criteria aimed at rejecting background events such as supernovae, we find the transient associated with NGC\,4993 as the only remaining plausible counterpart, and reject chance coincidence at the 99.5\% confidence level. We therefore conclude that the optical counterpart we have identified near NGC\,4993 is associated with GW170817. This discovery ushers in the era of multi-messenger astronomy with gravitational waves, and demonstrates the power of DECam to identify the optical counterparts of gravitational-wave sources.

\end{abstract}



\section{Introduction}
The joint detection of electromagnetic (EM) and gravitational wave (GW) emission from astrophysical sources is one of the holy grails of present-day astronomy. The primary targets for such joint detections are the mergers of compact object binaries composed of neutron stars (NS) and/or black holes.  In such systems the GW emission provides insight into the bulk motions, masses, binary properties, and potentially the composition of neutron stars.  Electromagnetic observations provide critical insights into the astrophysics of the event, such as the progenitor environment, the formation of relativistic and non-relativistic outflows, and in some cases the nature of merger products (e.g., \citealt{2012ApJ...746...48M,2013MNRAS.430.2585R,2017RPPh...80i6901B}). Combining EM and GW observations would lead to deeper  scientific insights into some of the most cataclysmic events in the Universe. These multi-messenger observations also allow for novel measurements, such as standard siren measurements of the Hubble constant \citep{1986Natur.323..310S,Holz2005,2006PhRvD..74f3006D,2010ApJ...725..496N,2013arXiv1307.2638N}, and studies of gamma-ray bursts \citep{2014ARA&A..52...43B}.

A wide range of EM emission mechanisms for GW sources have been proposed over the years \citep{2012ApJ...746...48M}, including short-duration GRBs \citep{1989Natur.340..126E,2007PhR...442..166N,2014ARA&A..52...43B}, on- or off-axis afterglow emission from radio to X-rays \citep{2011ApJ...733L..37V,2014MNRAS.445.3575C,2015ApJ...815..102F,2016ApJ...829..112L},  optical/near-IR emission due to  radioactive decay of $r$-process nuclei synthesized in the merger ejecta (so-called kilonova; \citealt{1998ApJ...507L..59L,1999A&A...341..499R,2010MNRAS.406.2650M,barnes2013}), and radio emission produced by interaction of the kilonova ejecta with the circumbinary medium \citep{2011Natur.478...82N,2012ApJ...746...48M}.

The search for optical counterparts is particularly attractive due to the combination of  emission that, unlike GRB emission, is not highly beamed and wide-field optical telescope facilities; a detection can then be followed up at other wavelengths with narrow-field instruments. Over the last two years, we have used the Dark Energy Camera \citep[DECam,][]{2015AJ....150..150F}, a 3 deg$^2$ wide-field imager  on the Blanco 4-m telescope at the Cerro Tololo Inter-American Observatory (CTIO), to follow up GW sources from Advanced LIGO \citep{2009RPPh...72g6901A} and Virgo \citep{2015CQGra..32b4001A} detectors \citep[see, e.g.,][]{2016ApJ...826L..13A,2016MNRAS.460.1270D}. In particular, we conducted rapid follow-up observations of the black hole binary merger events GW150914 \citep{2016PhRvL.116f1102A} and GW151226 \citep{2016PhRvL.116x1103A}, using DECam \citep{2016ApJ...823L..33S,2016ApJ...823L..34A,2016ApJ...826L..29C}.  No  optical counterpart was discovered in either case.

On 2017 August 17 at 12:41:06 UT the Advanced LIGO/Virgo (ALV) observatories detected a binary neutron star merger, GW170817 \citep[][]{GCN21505,GCN21509,LVC17}.  At  23:12:59 UT (10.53 hours after the GW detection) we began to image a 70.4 deg$^2$ region that covered 93\% of the localization probability in the map provided by the LVC at the time \citep{GCN21513}.
Immediately following the identification by one of us (R.~Chornock), we received a private communication from another DECam team member (R.~Foley) indicating that the source was also discovered in an image taken 0.5 hours ahead of ours by the Swope Telescope.
We issued a circular to the Gamma-ray Coordination Network (GCN) reporting the discovery at 01:15:01UT \citep{GCN21530}, including a reference to a GCN from the 1M2H collaboration at 01:05:23 UT (SSS17a; \citealt{GCN21529}), and subsequent to our GCN the DLT40 team also announced an independent detection (DLT17ck: \citealt{GCN21531} reported at 01:41:13 UT); see \citep{MMApaper} for an overview of the observations carried out by the community. This transient has received an International Astronomical Union name of AT2017gfo.

Subsequent to our discovery of the optical transient, we obtained follow-up observations with a wide range of telescopes, spanning radio to X-rays, which are detailed in the associated papers of this series: \citet{Cowperthwaite17,Nicholl17,Chornock17,Margutti17,Alexander17,Blanchard17,Fong17}. 

Here, in the first paper of the series, we present our DECam observations, the discovery of the optical transient, and a search for other potential counterparts across the 70.4 deg$^2$ region. We find no other potential optical
counterpart within the GW localization region, thus helping to significantly establish the association between the detected optical transient and GW170817. 
A measurement of the Hubble constant, the first utilizing a gravitational wave event as a standard siren measurement of distance \citep{Schutz1986,2006PhRvD..74f3006D}, is enabled by this work and is described in \citet{Hopaper}. 

\section{DECam Counterpart Search}

The alert for GW170817 was issued 40 minutes after the trigger, on 2017 August 17 at 13:21 UT \citep{GCN21505}, and was promptly received by our automated GCN listener system.  Two subsequent GCN circulars indicated that the high-significance candidate was consistent with a binary neutron star merger at $d\approx 40$ Mpc and coincident within 2 seconds with a short burst of gamma-rays detected by Fermi GBM \citep{GCN21505,GCN21509}.  Four hours later a sky localization map obtained from the three-detector ALV network was provided \citep{GCN21513}. 


The entire GW localization region was visible from Chile at the beginning of the night, setting within the first $\sim 1.5$ hours. Our DECam observations commenced at 23:13 UT (10.53 hours post merger) with 30 sec exposures in  $i$- and $z$-band.  
The resulting $5\sigma$ limiting magnitudes are $i\approx 22.0$ and $z\approx 21.3$ for point sources. The pre-determined sequence of observations consisted of 18 pointings (hexes), each with a 3 deg$^2$ coverage, with a second offset sequence to mitigate loss of area (e.g., due to gaps between CCDs).  The resulting areal coverage was 70.4 deg$^2$, 
corresponding to an integrated probability of 93.4\% of the initial GW sky map. Additional details of the pointing
and sequencing determination algorithm are available in~\cite{desgwCHEP}.
While our sequence of observations was on-going, a new localization map was released at 23:54 UT \citep{GCN21527}. While the overall shape of the two maps are similar, the probability peak was shifted significantly. 
In the revised map the integrated probability of our observations is 80.7\%. 

\subsection{Discovery and Observations}
\begin{figure*}[t!] 
\includegraphics[width=0.49\linewidth]{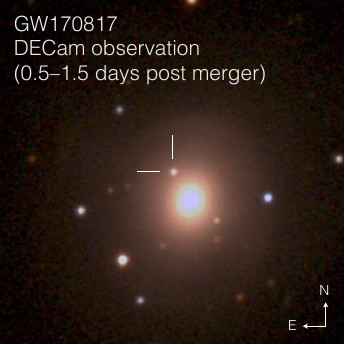}
\hfill
\includegraphics[width=0.49\linewidth]{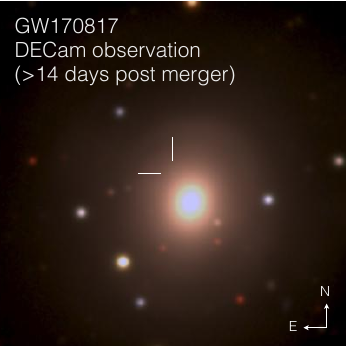}
\caption{NGC4993 $grz$ color composites (1.5'$\times$1.5'). Left: Composite of detection images, including the discovery $z$ image taken on 2017 August 18 00:05:23 UT and the g and r images taken 1 day later; the optical counterpart of GW170817 is at  RA,Dec $ = 197.450374, -23.381495$. Right: The same area two weeks later.}
\label{image}
\end{figure*}

\begin{figure}[h!]
\includegraphics[trim=230 40 210 40,clip, width=\linewidth]{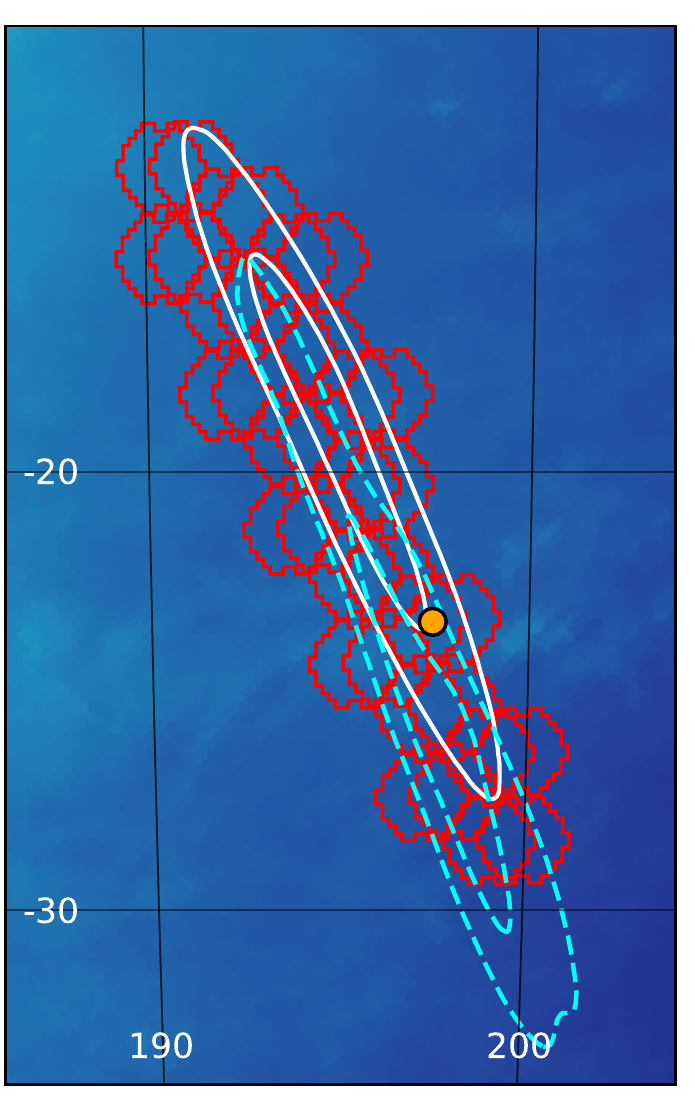}
\caption{Location of the optical counterpart of GW170817 on the probability maps provided by the LVC (white solid: initial; cyan dashed: revised; inner and outer contours show 50\% and 90\% probability respectively) and relative to our search area (red: DECam sky footprint).}
\label{map_with_source}
\end{figure}

We performed a visual inspection of raw, unprocessed DECam images to find new point sources near relatively bright galaxies in comparison to archival Pan-STARRS1 3$\pi$ survey images \citep{ps13pi}. This process resulted in the discovery 
of a new source near the galaxy NGC\,4993 (see Figure~\ref{image}).  The galaxy is located at $z=0.0098$ which is, for a value of $H_0$ of 70 km/s/Mpc, consistent with the $40 \pm 8$\, Mpc
reported by the LVC in their GCN for GW170817.  The transient is located at  coordinates RA,Dec $ = 197.450374, -23.381495$ 
(13h09m48.09s -23d22m53.38s) 
between the 50\% and 90\% contours in both the initial and shifted maps (see Figure~\ref{map_with_source}).

At the time when the galaxy was imaged (11.40 hours post-merger) the optical transient had magnitudes of
$i=17.30\pm 0.01$ and $z=17.43\pm 0.01$. 
We continued to observe the optical counterpart with DECam nightly in the $ugrizY$ filters until it became undetectable (at limiting magnitude $\sim$22.5 mag) in each band and the source location became inaccessible to the telescope.  Our last deep image of the source is on 2017 August 31, 14.5 days post-merger. These follow-up observations are discussed in detail in \citet{Cowperthwaite17}.

We process all images with the DES single-epoch processing~\citep[and references there in]{
2017arXiv170801531D,morganson17} and difference imaging (\texttt{diffimg}) pipelines \citep{2015AJ....150..172K}. The \texttt{diffimg} software works by comparing search images and one or more reference images (templates) obtained before or after the search images. We use our own imaging plus publicly available DECam data from the NOAO Science Archive (portal-nvo.noao.edu) as templates, requiring exposures of at least 30 sec. At the position of the counterpart, pre-existing templates were available in $g,r$ bands. For $u,i,z,Y$ images we used exposures taken after the source had faded ($u$: 25 August 2017; $i,z,Y$: 31 August 2017).

\begin{figure*}[t!]
\includegraphics[width=\linewidth]{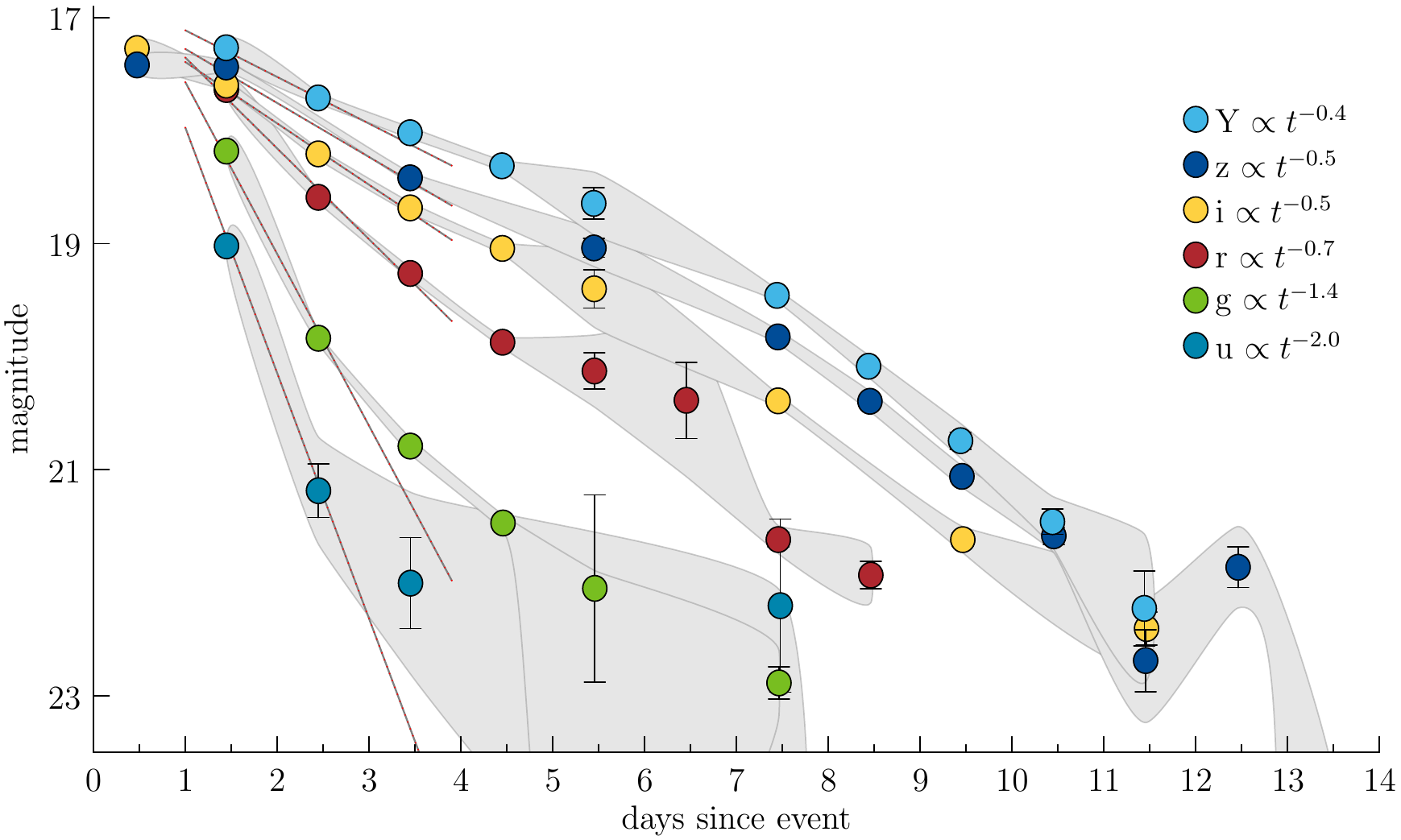}
\caption{
The observed light curve for the optical counterpart of GW170817 measured in 6 filters. The data
points and $1\sigma$ error bars were measured using \texttt{diffimg}. The gray bands represent the $95\%$ confidence interval about each measurement, or $2\sigma$ above sky if the measured flux is less than $2\sigma_{sky}$. Data taken on days 5.5, 6.5 and 11.5 were obtained in poor weather conditions, hence the broader uncertainty bands. 
The photometry has been checked against
a difference imaging reduction using Pan-STARRS PS1 templates 
\citep{Cowperthwaite17} and a galaxy fit \& 
subtract PSF source magnitude reduction; 
these measurements are consistent with
our results. 
The dashed lines show the initial decay of
the source: $m\propto t^{-\alpha}$, where $\alpha = (2.0,1.4,0.7,0.5,0.5,0.4)$ in u,g,r,i,z,Y bands respectively.
The decay in the first few days is consistent with a peak luminosity near $t = 1$ day. 
The table containing light curve information is available as a machine-readable file (Table~\ref{lct}).
\label{lc}}
\end{figure*}

The photometric results from \texttt{diffimg} are shown in Figure~\ref{lc} and Table~\ref{lct}.
The \texttt{diffimg} pipeline uses the well tested DES calibration module {\texttt{expCalib}}. The $ugrizY$ photometry presented in Table ~\ref{lct} has calibration errors relative to DES photometry of $\lessapprox 2\%$.
We implemented a galaxy morphological fit and subtraction method, making use of a fast multi-component fitting software \citep[\texttt{Imfit},][]{2015ApJ...799..226E} followed by PSF photometry 
and a Pan-STARRS PS1 calibration to double check the reduction. Results agree within uncertainties and calibration differences.  The
photometry used in the next paper in this series, 
\citep{Cowperthwaite17}, measured using a difference image reduction using Pan-STARRS PS1 templates, also agrees within uncertainties.

\begin{deluxetable}{rrrr}[hb!]
\tablecolumns{4}
\tablecaption{Light curve constructed from u,g,r,i,z,Y observations. Columns are observation time, band, magnitude and its errors. Magnitudes are galactic extinction corrected AB PSF magnitudes.
\label{lct}}
\tablehead{
    \colhead{MJD} & \colhead{Band} & 
    \colhead{Mag\tablenotemark{1}} & \colhead{$\sigma_m$}
    }
\startdata
57983.00306    &    i    &   17.27    &   0.005 \\
57983.00374    &    z    &   17.42    &   0.007 \\
57983.97395    &    Y    &   17.27    &   0.008 \\
\enddata
\tablenotetext{1}{E(B-V)$_{SFD98}$=0.123, and \\
R$_V$=3.963,3.186,2.140,1.569,1.196,1.048 
for u,g,r,i,z,Y bands, respectively.}
\tablecomments{Full table available as machine-readable in the online version. A portion is shown here for guidance regarding its form and content.}
\end{deluxetable}

\section{Wide Area Search}

Our primary program is to obtain images over the LIGO probability map to search for counterparts. 
Within 12 hours of the event we had obtained DECam $i,z$ images across $>80\%$ of the revised LIGO probability map. 
We have analyzed this region to determine how many potential counterparts are present.

\subsection{Image Processing}

We employ the DES single-epoch processing and \texttt{diffimg} pipelines to produce a list of transient candidates. The search images are 72 exposures taken on the night of the trigger (corresponding to two tilings of 18 hexes in $i$ and $z$ bands). Because most of the exposures ($\sim 60\%$) did not have pre-existing DECam imaging in those bands, we took images on the nights of 31 August through 2 September to serve as templates. We expect any kilonova-like transients to have faded below detection thresholds by that time ($\sim 2$ weeks from the trigger). 

The total area encompassed by the 72 exposures is 70.4 deg$^2$. 
The camera fill factor is 80\% 
which results in 4\% area loss when we consider the two overlapping tilings. Weather on the nights of template observations was partially clouded and caused a 3\% loss. Processing failures accounted for less than 1\% loss. The final area included in this search for other potential counterparts is 64.6 square degrees.

This process results in 1500 transient candidates with magnitudes between 15.5 and 20.5. A candidate is defined as a detection meeting \texttt{diffimg} quality requirements (see Table 3 of~\cite{2015AJ....150..172K}) on at least two search exposures. 
The magnitude cutoff of this analysis is limited by the depth of the template images: $i=21.2$ and $z=20.5$, to be compared with the depth of the search images: $i=22.0$ and $z=21.3$.

\subsection{Candidate Selection}

We apply several selection criteria to candidates identified in \texttt{diffimg} aiming to reject moving objects, background artifacts and long-lived transients:
\begin{itemize}

\item Criterion 1: The candidate must have at least one detection in $i$ and one detection in $z$ band. 

\item Criterion 2: The candidate must pass our 
automated scanning program \citep{2015AJ....150...82G} with a machine learning score $\geq 0.7$ in all detections. This criterion rejects non-astrophysical artifacts. The efficiency of this criterion as measured from point sources injected into our images is $\approx 90-100\%$ for both $i$ and $z\approx 17-22$ mag.

\item Criterion 3: The candidate must have faded significantly between the search (first) and template (last) observations.  Specifically, we require the change in the candidate flux to be greater, at 3$\sigma$ level, than the flux in a circular aperture of radius = 5 pixels on the template image.
image at the location of the candidate. This criterion eliminates slow-evolving transients (e.g., supernovae). 
\end{itemize}

Table~\ref{cuts} lists the number of events passing each selection stage in various bins of magnitude. After all criteria are applied, 
one optical counterpart candidate remains: the source discovered by visual inspection. 
Slow moving solar system objects, which could potentially have met the
selection criteria above, are very rare in the magnitude range of this search.
The flares of M dwarf flare stars have $T\approx $10,000\,K and therefore are very blue; they are rejected by selection criterion~3 in $z$-band.

\begin{deluxetable}{lrrrr}[h!]
\tablecolumns{5}
\tablecaption{Number of candidates at each selection stage, sorted by $i$-band magnitude.  
\label{cuts}}
\tablehead{
    \colhead{mag($i$)} &
    \colhead{Raw} & \colhead{Cut 1} &
    \colhead{Cut 2} & \colhead{Cut 3} }
\startdata
15.5--16.5 &    4 &   0 &  0 & 0 \\
16.5--17.5 &   11 &   7 &  3 & 1 \\
17.5--18.5 &   26 &  15 &  7 & 0 \\
18.5--19.5 &  296 &  63 & 27 & 0 \\
19.5--20.5 & 1163 & 167 & 44 & 0 \\
\tableline\\[-15pt]
Total      & 1500 & 252 & 81 & 1 \\ 
\enddata
\end{deluxetable}

\section{Uniqueness of the Candidate}

This analysis shows that the source we discovered is the only one plausibly associated with the GW event within the region searched.  To estimate its significance we compute the chance probability of a transient to occur within the volume and timescale of interest. Because SNe are by far the most likely transient contaminant, we use their rate and timescale to make a conservative estimate. We use a combined rate of $1 \times 10^{-4}$ Mpc$^{-3}$yr$^{-1}$, for core-collapse \citep{2015ApJ...813...93S} and Type Ia \citep{2008ApJ...682..262D} SNe at $z\lesssim 0.1$.  The characteristic timescale of SNe is $\tau \sim $1 month.  The volume ($V$) we observed is estimated as a shell at $z\sim 0.01$ ($\sim 40$ Mpc) spanning 64 deg$^2$ area and 16 Mpc width corresponding to the effective search area and the distance uncertainty, respectively: $V=558$ Mpc$^3$. Under these assumptions, we find that the probability of a chance coincidence is $\sim 0.5\%$, and we conclude that our optical transient is associated with GW170817.

\section{Conclusion}

We report the DECam discovery of the optical counterpart to the BNS merger GW170817, an object with $i= 17.30$ mag and $z= 17.43$ mag at 11.40 hours post-merger. 
The source was discovered through visual inspection of nearby galaxies in our raw data stream.
Our analysis identifies this source as the  only credible optical counterpart within a large fraction of the GW170817 sky map.
The observed peak absolute magnitude of $M_i=-15.7$ is about 1000 times brighter than a nova, which is typically close to Eddington luminosity  ($M_V=-9$). 
Thus, we have indeed discovered a kilonova as the name defines it and was  predicted in \cite{2010MNRAS.406.2650M}.

At $M_i = -15.7$, the optical transient is bright enough for us to detect it out to 425 Mpc. Its properties, 1.5 days after the event include: $(i-z)=0.2$, and a magnitude decline versus time in $i,z,Y \propto t^{-\frac{1}{2}}$, and faster decline in $g$-band ($\propto t^{-\frac{3}{2}}$).
Future searches for counterparts of GW events may be improved by this information.

This detection has opened a new era of multi-probe multi-messenger astronomical observations of the Universe that will bring new measurements of cosmological parameters, starting with the present rate of expansion~\citep{Hopaper}, and possibly helping determine the matter/energy content and evolution of the Universe.

This is the first detection of an optical counterpart of a gravitational wave source. It will not be the last. 
As the LIGO and Virgo collaborations proceed to their next observing runs and upgrades, DECam will continue to play an important, almost unique, role in the identification of gravitational wave sources in the Southern hemisphere.

\acknowledgments
Funding for the DES Projects has been provided by the DOE and NSF(USA), MEC/MICINN/MINECO(Spain), STFC(UK), HEFCE(UK). NCSA(UIUC), KICP(U. Chicago), CCAPP(Ohio State), 
MIFPA(Texas A\&M), CNPQ, FAPERJ, FINEP (Brazil), DFG(Germany) and the Collaborating Institutions in the Dark Energy Survey.

The Collaborating Institutions are Argonne Lab, UC Santa Cruz, University of Cambridge, CIEMAT-Madrid, University of Chicago, University College London, 
DES-Brazil Consortium, University of Edinburgh, ETH Z{\"u}rich, Fermilab, University of Illinois, ICE (IEEC-CSIC), IFAE Barcelona, Lawrence Berkeley Lab, 
LMU M{\"u}nchen and the associated Excellence Cluster Universe, University of Michigan, NOAO, University of Nottingham, Ohio State University, University of 
Pennsylvania, University of Portsmouth, SLAC National Lab, Stanford University, University of Sussex, Texas A\&M University, and the OzDES Membership Consortium.

Based in part on observations at Cerro Tololo Inter-American Observatory, National Optical Astronomy Observatory, which is operated by the Association of 
Universities for Research in Astronomy (AURA) under a cooperative agreement with the National Science Foundation.

The DES Data Management System is supported by the NSF under Grant Numbers AST-1138766 and AST-1536171. The DES participants from Spanish institutions are partially 
supported by MINECO under grants AYA2015-71825, ESP2015-88861, FPA2015-68048, and Centro de Excelencia SEV-2012-0234, SEV-2016-0597 and MDM-2015-0509. Research leading 
to these results has received funding from the ERC under the EU's 7$^{\rm th}$ Framework Programme including grants ERC 240672, 291329 and 306478.
We acknowledge support from the Australian Research Council Centre of Excellence for All-sky Astrophysics (CAASTRO), through project number CE110001020.

This manuscript has been authored by Fermi Research Alliance, LLC under Contract No. DE-AC02-07CH11359 with the U.S. Department of Energy, Office of Science, Office of High Energy Physics. The United States Government retains and the publisher, by accepting the article for publication, acknowledges that the United States Government retains a non-exclusive, paid-up, irrevocable, world-wide license to publish or reproduce the published form of this manuscript, or allow others to do so, for United States Government purposes.

HYC, ZD, BF, MF, and DEH were partially supported by NSF CAREER grant PHY-1151836 and NSF grant PHYS-1708081. They were also supported by the Kavli Institute for Cosmological Physics at the University of Chicago through NSF grant PHY-1125897 and an endowment from the Kavli Foundation. 

The Berger Time-Domain Group at Harvard is supported in part by the NSF through grants AST-1411763 and AST-1714498, and by NASA through grants NNX15AE50G and NNX16AC22G.

We thank the University of Copenhagen, DARK Cosmology Centre, and the Niels Bohr International Academy for hosting R.J.F., Z.D., B.F.\ during the discovery of GW170817/SSS17a, where they were participating in the Kavli Summer Program in Astrophysics, "Astrophysics with gravitational wave detections". This program was supported by the Kavli Foundation, Danish National Research Foundation, the Niels Bohr International Academy, and the DARK Cosmology Centre.

The UCSC group is supported in part by NSF grant AST--1518052, the Gordon \& Betty Moore Foundation, the Heising-Simons Foundation, generous donations from many individuals through a UCSC Giving Day grant, and from fellowships from the Alfred P.\ Sloan Foundation and the David and Lucile Packard Foundation to R.J.F.

We thank the CTIO director, Stephen Heathcote,
for being supportive of our target of opportunity program using DECam. We also thank the CTIO staff for providing a great experience to observers on-site and remotely.

\bibliographystyle{yahapj}
\bibliography{references}

\end{document}